\RequirePackage{fix-cm}
\documentclass[smallextended]{svjour3}       
\smartqed  
\usepackage{graphicx}
\begin{document}

\title{On the uniqueness of the space-time energy in General Relativity. The illuminating case of the Schwarzschild metric
}

\titlerunning{On the uniqueness of the space-time energy. The case of the Schwarzschild metric}        

\author{Ramon Lapiedra \and Juan Antonio Morales-Lladosa}


%
\institute{Ramon Lapiedra \and Juan Antonio Morales--Lladosa \at Departament
d'Astronomia i Astrof\'{\i}sica,
\\Universitat
de Val\`encia, E-46100 Burjassot, Val\`encia, Spain.\\
Tel.: +34-96-3543066\\
Fax: +34-96-3543084\\
\email{ramon.lapiedra@uv.es}\\
\email{antonio.morales@uv.es}}

\date{Received: date / Accepted: date}

\maketitle

\begin{abstract}
The case of asymptotic Minkowskian space-times is considered. A special class of asymptotic rectilinear coordinates at the spatial infinity, related to a specific system of free falling observers, is chosen. This choice is applied in particular to the Schwarzschild metric, obtaining a vanishing energy for this space-time. This result is compared with the result of some known theorems on the uniqueness of the energy of any asymptotic Minkowskian space, showing that there is no contradiction between both results, the differences becoming from the use of coordinates with different operational meanings. The suitability of Gauss coordinates when defining an {\em intrinsic} energy is considered and it is finally concluded that a Schwarzschild metric is a particular case of space-times with vanishing {\em intrinsic} $4$-momenta.

\end{abstract}



\section{Introduction}
\label{intro}

It has been largely discussed and carefully established  that there is a sound definition of energy (and also of linear $3$-momentum and angular $4$-momentum) of any space-time which is asymptotically Minkowskian, once we have selected a symmetric complex, as for example, the one of Weinberg \cite{Weinberg} or the other one of Landau and Lifshitz \cite{Landau}, or some other rather different, but likely equivalent,  prescriptions (cf. \cite{ADM-1961,Ashtekar,Katz-Bi-LyBell-97,Chang-Nester-Chen-1999} for instance).

Whatever it be, it is generally assumed (see \cite{Weinberg} for example) that in order to obtain a sound definition of this energy we must rely on some coordinate system which, fast enough, becomes a rectilinear one in the spatial infinity and then, to simplify the calculation, use the 
Gauss theorem to write the energy $3$-volume integral as a $2$-surface integral at this infinity 
(see \cite{EricG,Alcubierre} for a clear account on this and related topics).  
Nevertheless, this theorem can only be applied if the first derivatives of the field in the $3$-volume integrand are continuous, though in some situations we can overcome these conditions by redefining, in the sense of distributions, non smooth enough derivatives, as it is for instance the case for an elementary charge in electrostatics. We will take the fact of these conditions into account along the present paper.

Furthermore, is this sound definition of energy unique? According to some well known theorems \cite{Brill-Deser,Schoen-Yau,Witten,Parker-Taubes}, the answer is ``yes", provided that the metric approaches asymptotically the Minkowski metric fast enough, but not too fast (see, to begin with, the readable considerations presented in  \cite{Weinberg} and our Appendix \ref{ap-A} for some comments on it). We will consider two cases of these different asymptotic approaches,  first a non-rotating, non-charged, black hole, and then a Schwarzschild metric 
with a source radius larger than the Schwarzschild radius.

All the same, a precision is needed here since these uniqueness theorems start from the Arnowitt-Deser-Misner (ADM)  definition of the $P^0$ energy (see \cite{ADM}), or what can be  equivalent, to this regard, from the Weinberg complex. Then, if we change to another of the well known complexes in the literature, the above uniqueness can not remain true. For instance, from \cite{AguirreAV-96} one concludes that for a Schwarzschild metric in ``Cartesian" Kerr-Schild coordinates, the use of Landau-Lifshitz complex leads to the well result $P^0 = m$, while the use of the same complex in  ``Cartesian" static coordinates \cite{Vir-91} leads to an everywhere negative energy density.

Next,  we will comment on the alternative to  these uniqueness theorems and finally discuss what particular energy can be considered what will be called the {\em intrinsic} energy of this black hole or this Schwarzschild metric with this larger radius.

Let us recall the expression for the energy, $P^0$,  of an asymptotically Min\-kows\-kian metric, $g_{\alpha\beta}$,  on the Weinberg complex basis \cite{Weinberg}:

\begin{equation} \label{energy}
P^0 =  \frac{1}{16 \pi} \int \partial_i (\partial_j g_{ij} - \partial_i g_{jj}) dx^{1} dx^{2} dx^{3} 
\end{equation}
where $i,j, ... =1,2,3$, stand for the $3$-space indices, $g_{jj} \equiv \delta_{ij} g_{ij}$, $\partial_j g_{ij} \equiv \delta_{jk} \partial_j g_{ik}$, 
and we have taken the gravitational constant, $G$, and the speed of light, $c$, equal to $1$.
Notice that the Weinberg procedure leads to the same energy expression (\ref{energy}) 
that the one proposed previously by Arnowitt, Deser and Misner  (in the form of next Eq. (\ref{energy-S})) following a Hamiltonian approach \cite{ADM} for the Einstein field equations.

Let us now  consider the particular case of the Schwarzschild metric in the static standard coordinates:
\begin{equation}\label{Smetrica}
 ds^2= - (1 - \frac{r_0}{r})dt^2 +  \frac{dr^2}{1 - \frac{r_0}{r}} + r^2 d \sigma^2, \quad  r_0\equiv 2m, 
\end{equation}
where $m$ is the source mass,  and $d \sigma^2 \equiv d \theta^2 + \sin^2 {\hspace{-0.7mm}} \theta \, d \phi^2$ is the metric on the unit $2$-sphere. The metric (\ref{Smetrica})
is obviously asymptotically Minkowskian for $r \to  \infty$,  going like $r^{-1}$ at this distant infinity, and it has a non intrinsic 
singularity for $r = r_0$, and another intrinsic one for $r=0$.

If the radius of the spherical source is larger than $r_0$, this singularity at $r=r_0$ is not more present, 
and because of the Jebsen-Birkhoff theorem \cite{Jebsen-Birkhoff}, the same happens for the intrinsic singularity at $r=0$ (letting aside particular cases as the one of a 
black hole surrounded by a spherical non rotating shell at some large radius \cite{Shell-BH}). Then using Weinberg complex, for example, 
we can apply Gauss theorem and write (\ref{energy}) as the $2$-surface integral  on the boundary $r = \infty$,  
\begin{equation} \label{energy-S}
P^0 =  \frac{1}{16 \pi} \lim_ {r \to \infty}  \int (\partial_j g_{ij} - \partial_i g_{jj}) n_i r^2 d\Omega, 
\end{equation}
where as in (\ref{energy}) the contractions in the $3$-space indices are performed  with the Kronecker  $\delta_{ij}$, and $n_i = \delta_{ij} x^j/r$, $r^2 = \delta_{ij} x^{i} x^{j}$, 
$d \Omega = \sin \theta d \theta d\phi$.
As it is well known, applying (\ref{energy-S}) to the metric (\ref{Smetrica}) in rectilinear coordinates gives the value $P^0 = m$, provided that, as mentioned above, the radius of the spherical source is larger than $r_0$ (see \cite{Weinberg} for example).

But, what is the value of $P^0$ when (\ref{Smetrica}) represents a black hole so that both metric singularities, at $r=r_0$ and $r=0$, remain present? 
To begin with, because of the singularity at $r = r_0$, Gauss theorem cannot be applied to the $3$-volume integral in (\ref{energy}), 
while the partial contribution to this integral, from $r = \infty$ to $r \to r_0$, can be easily seen to diverge in this limit $r \to r_0$.

To overcome this difficulty, we will calculate the {\em intrinsic} energy $P^0$ for a black hole, using a convenient family of coordinate systems. 
We will obtain $P^0 = 0$ (Sect. \ref{sec-2}), and then we will comment this result (Sect. \ref{sec-3}) and compare it with the other well known result on the subject, $P^0 = m$, 
when the source radius is larger than $r_0$ (Sect. \ref{sec-4}). The remaining sections of the paper are devoted to discuss what could be considered 
as the {\em intrinsic} energy of a gravitational field with its sources (Sect. \ref{sec-5}), and then to apply this notion to establish the {\em creativeness} of the Schwarzschild geometry 
(Sect. \ref{sec-6}) and, finally, to justify the goodness of the Weinberg complex in the present issue  (Sect. \ref{sec-7}). Appendices \ref{ap-A} and  \ref{ap-B} contain 
detailed proofs of important results and considerations  used to achieve the main results. 

A summary of the results of this work has been recently presented at the Spanish Relativity Meeting in Portugal-ERE2012 \cite{ERE-2012}.

%
%
%


\section{Calculating the {\em intrinsic} energy of a black hole}
\label{sec-2}

Let us consider the metric of a non rotating, non charged black hole, referred to Lema\^{i}tre coordinates, 
\begin{equation}\label{Lemaitre}
 ds^2 = - dT^2 +  \frac{r_0}{r} dR^2+ r^2 d \sigma^2, 
\end{equation}
(see  \cite{Landau}, epigraph $102$, or \cite{Elmestre})  where the function $r$ defined as 
\begin{equation} \label{r(RT)}
r^{3/2} \equiv k (R - \eta T), \quad k\equiv \frac{3}{2} \sqrt{r_0}, \quad \eta \equiv \pm 1, 
\end{equation}
takes all values between $0$ and $ \infty$. The Kruskal-Szekeres black and white regions are respectively described with the two coordinate branches provided by 
$\eta = 1$ and $\eta = -1$ (see \cite{Schutzhold}). The following considerations apply equally to both (the black and the white) regions, although we shall only refer 
explicitly to the first one.

We can obtain (\ref{Lemaitre}) by making the following change of 
the $t$ coordinate 
\begin{equation} \label{canvi}
T = t + \eta \, r_0 f(r),  \quad f(r) \equiv 2 \sqrt{\frac{r}{r_0}} + \ln \Big|\frac{\sqrt{r} - \sqrt{r_0}}{\sqrt{r} + \sqrt{r_0}}\Big|, 
\end{equation}
in (\ref{Smetrica}), taking (\ref{r(RT)}) into account. This time dependent metric  (\ref{Lemaitre}) is regular everywhere except for the intrinsic singularity $R - \eta T = 0$, 
corresponding to $r = 0$. 

Furthermore, it is written in Gauss coordinates, the metric component $g_{0\alpha}$ being $g_{00}=-1$, $g_{0i}=0$. This implies that the curves $R = R_0$, 
$\theta = \theta_0$, $\phi = \phi_0$, with $R_0$, $\theta_0$ and  $\phi_0$ constants, are time-like geodesics of the metric. In 
other words, they describe free-falling particles.

As largely explained in \cite{NewCreatable} this kind of coordinates is the only one  that has to be considered 
to define an {\em intrinsic} energy and momenta of a space-time in General Relativity,  named {\em proper} energy and momenta in \cite{NewCreatable}. 
We will come back next (see Sect. \ref{sec-4}) to the question and to the change of name.

But metric (\ref{Lemaitre}) does not approach the Minkowski space-time, $M_4$,  for $R \to \infty$ at fixed $T$, or which is equivalent for 
$r \to \infty$. This means that the present coordinates 
$(T, R)$ are not good coordinates  in order to calculate $P^0$. Then, to reach some good ones, let us change the $R$ 
coordinate and go to the new coordinate $\rho$ defined as
\begin{equation} \label{rhoR}
\rho^{3/2} + C =  k R , \quad \rho > 0, 
\end{equation}
with $C$ some arbitrary constant.

Then, in the new coordinates $(T, \rho, \theta, \phi)$,  the metric (\ref{Lemaitre}) becomes:
\begin{equation}\label{LemaitreP}
ds^2  =  - dT^2 +  \frac{\rho}{r} d\rho^2+ r^2 d \sigma^2 \equiv  - dT^2 +   dl^2, 
\end{equation}
where now $r$ can be written:
\begin{equation} \label{rrhoT}
r^{3/2} = \rho^{3/2} - \eta k T + C. 
\end{equation}

Notice that the $3$-metric  $dl^2 = \rho/r d\rho^2+ r^2 d \sigma^2$ is flat (its Ricci tensor vanishes) and further, for any given value of $T = T_0$ we can select a $C$ value such that $C = \eta k T_0$,  and so such that $r = \rho$, for $T = T_0$, which means that, for any $T_0$, we can select a corresponding coordinate  system where
\begin{equation}\label{LemaitreT} 
 ds^2|_{T_0} \equiv ds^2(T = T_0) =  - dT^2 +  d\rho^2+ \rho^2 d \sigma^2, 
\end{equation}
that is the Minkowski metric everywhere on the $3$-surface $T=T_0$,  up to for the essential singularity 
$r=0$ ($R = \eta T_0$), written in spherical coordinates. As a consequence, for each selected  $T_0$,  
$P^0$ vanishes,
\begin{equation}\label{Pzero}
P^0 = 0, 
\end{equation}
according to the expression (\ref{energy}), since in rectilinear $x^{i}$ coordinates,  $dl^2$ defined in (\ref{LemaitreP}) becomes 
$dl^2 = \delta_{ij} x^{i}x^{j}$ and so the same energy density vanishes everywhere except for $\rho = 0$, this exception being by no means always irrelevant, according to 
the discussion contained in the following section.


\section{A discussion on the vanishing energy and the {\em more than quasi-local} algorithm}
\label{sec-3}

A similar result to the above  vanishing energy density is reported in \cite{AguirreAV-96,Vir-91,Vir-90ab} for a quite large family of metrics containing the Schwarzschild one, dealing with any of the Einstein, Tolman, Landau-Lifshitz, M{\o}ller, Papapetrou and Weinberg complex, and using some non static ``Cartesian coordinates" approaching Minkowski metric at the 3-space infinity. 

Our $P^0 = 0$ value is more precisely defined, first by excluding from the entire volume in  (\ref{energy}) a small ball of radius $\varepsilon$ covering the singularity $\rho = 0$, 
and then by taking the limit for $\varepsilon \to 0$. From (\ref{LemaitreT}) this result, $P^0 = 0$, is obviously true irrespective of the complex used, the one from 
Weinberg or any other one. Further, as explained in the penultimate paragraph of the following section, we find the result consistent with the one we will obtain, 
in Sect. \ref{sec-4},  for $P^0$ in the case where the Schwarzschild source radius is larger than the corresponding Schwarzschild radius, in the above coordinates $\{T, \rho, \theta, \phi\}$.

Notice, all the same, the kind of  algorithm used to define $P^0$ in the black-hole case: we 
first select a given space-like $3$-surface $\Sigma_3(T_0)$, defined as $T=T_0$, then,  by selecting a suitable value of the constant $C$ in (\ref{rrhoT}), we use a particular coordinate system $(T, \rho, \theta, \phi)$ so that, for $T=T_0$, (\ref{LemaitreT}) is satisfied, and finally we calculate $P^0$ for $\Sigma_3(T_0)$ in this suitable coordinate system.  As explained,  the calculation gives trivially $P^0 = 0$, independently of $T_0$, provided that for every new $T_0$ value we suitably change the constant value $C$. We call this algorithm the {\em more than quasi-local} algorithm because  at the very beginning it involves an integration on a whole $3$-volume instead on its boundary and we say next something more on it.

But, why have we not here followed the standard procedure, that is, first calculating $P^0$ for any $T$ and then particularizing for 
$T = T_0$?  The reason is that (\ref{LemaitreP})  presents a physical singularity at $r=0$. Then, a $3$-volume integral like  
 (\ref{energy}) has to be defined first by subtracting an elementary spherical $3$-volume, $r = |\epsilon|$, and then taking 
 the limit $|\epsilon| \to 0$. Since no other singularity is present in (\ref{LemaitreP}), we can apply the Gauss 
 theorem to this truncated $3$-volume integral, and express it like the corresponding flux through the infinite boundary $\rho \to r \to  \infty$, 
 plus the contrary flux through $r = |\epsilon|$, before taking finally the limit $|\epsilon| \to 0$. The problem with this hypothetical calculation 
 is that while (\ref{LemaitreP}) is asymptotically Minkowskian for $\rho \to  \infty$, it is not asymptotically Minkowskian for $r =  (\rho^{3/2} - \eta k T + C)^{2/3} \to 0$. 

 As far as the first limit, $\rho \to \infty$, is concerned, it can be easily seen from (\ref{rrhoT}) that metric (\ref{LemaitreP}) goes 
 like $ \rho^{-3/2}$ towards Minkowski metric and so faster than $\rho^{-1}$ which is the limit law of decreasing to make sure 
 the convergence of $P^0$ defined in (\ref{energy-S}). Let us remark that a  proper definition of asymptotically flatness (see for example \cite{EricG}) should include the suitable kind of asymptotic behavior of the time derivatives of the metric. In our case, we do not need to consider the behavior of these time derivatives since they do not appear neither in
(\ref{energy})  nor in (\ref{energy-S}). They do appear in the corresponding integral expressions of the linear $3$-momentum, $P^{i}$, and the angular $4$-momentum, 
$J^{\alpha \beta}$, $\alpha, \beta =0, 1, 2, 3$, of the space-time considered \cite{Weinberg}. But $P^{i}$ and  
$J^{\alpha \beta}$ vanish in our case (the Schwarzschild metric) irrespective of the kind of asymptotic behavior, because of the present spherical symmetry.

As far as the second limit is concerned, we cannot use asymptotic rectilinear coordinates at the boundary $|\epsilon| \to 0$. To circumvent this difficulty, we could try to generalize 
 this Minkowskian prescription demanding that in a suitable new  Gauss coordinate system the new  $dl'^2$ (see (\ref{LemaitreP})) becomes  
 manifestly conformally flat for $|\epsilon| \to 0$ (in accordance with the ideas exposed in \cite{NewCreatable} for {\em universes} which are non asymptotically Minkowskian).
 But this is not possible, not even for an elementary neighborhood of $T=T_0$,  because it can be seen (Appendix \ref{ap-B}) that  the only solution of the vacuum Einstein field equations like 
 \begin{equation}\label{Eq-prima} 
 ds^2 =   - dT'^2 +  G(T', \rho') (d\rho'^2+ \rho'^2 d \sigma^2)
\end{equation}
with $G(T', \rho')$ any regular function of $T'$ and  $\rho'$, is just (locally) the Minkowski space-time $M_4$. 

Of course, we could loose this coordinate condition by using Gauss coordinates such  that we have (\ref{Eq-prima}) only over the boundary 
$2$-surface  of the space-like  $3$-surface $T = T_0$ (this can always be done: see again \cite{NewCreatable} and references therein).
But this is less than what we have already reached: to have (\ref{LemaitreT}), that is (\ref{Eq-prima}) with $G(T', \rho') = 1$ everywhere on $T=T_0$.  Therefore, the coordinates used to write the 
Schwarzschild metric in the form (\ref{LemaitreP})  are the good (or at least simple good) coordinates to define the {\em intrinsic} energy $P^0$ of our 
 black hole, using the so called {\em more than quasi-local} algorithm, which becomes $P^0 = 0$. 

 We discuss in the next section in which sense we can call this vanishing energy of the black hole its {\em intrinsic} energy, and we will compare it with the different result $P^0 = m$  for the Schwarzschild metric, concluding that there is no contradiction between these two different results.

 But,  before leaving the present section, let us say some words about what we have called above ``the more than quasi-local" algorithm. 
 Remember that we have been not able to define first $P^0$ for any time $T$, in a consistent way (i.e.,using Gauss coordinates which behave appropriately at the suitable boundary),  
 in order to afterwards particularize it for $T=T_0$. Because of this, 
 the algorithm renounces to associate an energy to a time--dependent metric in General Relativity, the energy remaining then associated to any 
 space-like $3$-surface, $\Sigma_3(T_0)$, which is less restrictive than the {\em quasi-local} energy program (for a review, see \cite{Szabados-review}), where the energy $P^0$ is namely associated to any closed $2$-surface embedded in $\Sigma_3(T_0)$. These considerations explain why we have referred to our algorithm as to a more than quasi-local one.

 But the algorithm has the supplementary virtue of having led us to a result, $P^0 = 0$,  for an Schwarzschild black hole in  Gauss coordinates, which was to be expected. To see it, notice that the 
 same result for the same kind of coordinates will be obtained (Section \ref{sec-4}) for a Schwarzschild metric whose source radius is larger than $r_0$ (a suitable ideal star), this time directly, we 
mean without using the algorithm. But we could imagine that our ideal star has enough mass as to undergo an ideal collapse preserving the spherical symmetry and without expelling any mass 
(of course, without  radiating any gravitational energy too). In this process, we can hope that $P^0$ has to remain constant. Thus if initially was $P^0 = 0$, this should 
 be the remaining value when the collapse has been completed, which is just the result obtained using our algorithm.

 Explaining it in more general terms, let us accept that our isolated already collapsed black hole has to be a physical system independent of the origin of the physical time $T$ of (\ref{LemaitreP}).  In other words, it is always the same time--dependent physical system. In particular, its energy should exhibit no dependence on this origin. Thus, we can select any time $T_0$ to calculate it. But, this is just what does the {\em more than quasi-local} algorithm, with the final result that, for (\ref{LemaitreP}), the corresponding $P^0$ does not depend on the free selected $T_0$, giving precisely $P^0 = 0$.
%


\section{Vanishing energy versus the mass energy for a Schwarzschild metric}
\label{sec-4}

The energy, $P^0$, associated to the Schwarzschild metric, (\ref{Smetrica}), 
in static standard coordinates,  takes the value $P^0 = m$, provided that we use the Weinberg complex, for example, and provided that 
the radius  of the spherical source is larger than $r_0$ \cite{Weinberg}. The  same result is obtained using another complexes and another 
asymptotic rectilinear coordinates in \cite{AguirreAV-96,Vir-90ab,Virbhadra}. See also \cite{Xulu-tesi} for a readable account on the subject.

Then, let us consider the Schwarzschild metric in the form (\ref{LemaitreP}) when again the source radius is larger than $r_0$. This form of the metric has no 
singularities, and we can apply Gauss theorem, so as to arrive to (\ref{energy-S}).

But it is very easy to see that, for any given $T$ value, metric (\ref{LemaitreP}) approaches like $\rho^{-3/2} \approx r^{-3/2}$ the $M_4$ metric 
at the spatial infinity $\rho \to  \infty$.

From (\ref{energy-S}) and this kind of approaching law it is easy to see with rather no calculation that metric (\ref{LemaitreP}) gives the value $P^0 = 0$, the same value, by the way, 
that the one obtained  for this metric with a vanishing source radius in the precedent section, using the {\em more than quasi-local} algorithm. 

This result, $P^0 = 0$, could seem erroneous at a first view since there is a well known statement (cf. \cite{Brill-Deser,Schoen-Yau,Witten,Parker-Taubes}), for asymptotically Minkowskian spaces, stating, under wide hypothesis including the non-negative character of the matter energy density, that the only space among these spaces having $P^0 = 0$ is the Minkowski space. But there is no contradiction between our result, $P^0 = 0$, and this theorem, 
since this one assumes, in particular,  that the $3$-space metric components, $g_{ij}$, go to $\delta_{ij}$ like $r^{-1}$ when 
$r \to \infty$, while in our case we have $g_{ij} = \delta_{ij} + O(r^{-3/2})$. 

Notice, then, that our result does not mean at all that this theorem, like same sharper ones \cite{Murchadha,Bartnik,Chrusciel,Beig-Chrusciel} in the literature, were wrong, since we only have been able to circumvent their correct results by changing its plausible boundary conditions, $g_{ij} -\delta_{ij} = O(r^{-1})$ or even $o(r^{-1/2})$,  by other different ones, 
$g_{ij} -\delta_{ij} = O(r^{-3/2})$, unavoidable in a different context: the last boundary conditions are not some ones selected {\em ad hoc}  in order to obtain a vanishing space-time energy, on the contrary, they come from the physical requirement that in order to define an {\em intrinsic} energy we should use coordinate systems whose coordinate time was a 
physical and {\em universal} time, that is, we should use Gauss coordinates, at least in the elementary neighborhood at the $r \to \infty$ boundary (see next section). In other words, 
as it was stressed in Ref. \cite{Zannias},  our metric falloff,  $g_{ij} = \delta_{ij} + O(r^{-3/2})$, and  the corresponding vanishing of $P^0$ are not necessarily in contradiction with the results stablished in the aforementioned references   \cite{Murchadha,Bartnik,Chrusciel,Beig-Chrusciel}.

In any case, the fact we want to stress here is that we do not obtain the same value for $P^0$ for the same space-time 
(a Schwarzschild metric whose  source radius is larger than $r_0$) when we use static standard coordinates (see Eq. (\ref{Smetrica})), in which case we obtain $P^0 = m$, than when we use the coordinates $(T, \rho, \theta, \phi)$ of (\ref{LemaitreP}), in which case we obtain $P^0 = 0$, though in both cases the metric becomes asymptotically 
Minkowskian fast enough: like $r^{-1}$ in the first case, and like $r^{-3/2}$ in the second case. Notice that, in any case, this result would not be invalidated by  Weinberg's statement in \cite{Weinberg} (see epigraph 6, chapter 7), according to which when an asymptotically Minkowskian space-time is referred to different coordinate systems, 
all them manifestly Minkowskian at the spatial infinity, we always found the same Minkowskian $4$-vector for the linear $4$-momentum. It would not be invalidated because, when trying to prove his statement, Weinberg assumes
implicitly that both approaches are as fast as $r^{-1}$, while in our case only one of the two approaches is of this kind, the other going like  $r^{-3/2}$ (see Appendix \ref{ap-A}).


\section{Intrinsic energy and Gauss coordinates}
\label{sec-5}

As it has been already commented in the Introduction, a sound definition of the energy, $P^0$, of an asymptotically Minkowskian space-time, must rely on
coordinate systems which be rectilinear at the spatial infinity $r \to  \infty$. Nevertheless, we have just seen that even in those kind of coordinate systems, a Schwarzschild metric whose source radius is longer than 
$r_0$, has different values for $P^0$ (one of them vanishing) when using different coordinate systems all them becoming fast enough rectilinear at this infinity. Thus, which 
if any of these boundary rectilinear coordinate systems should be chosen in order to calculate the sound physical energy $P^0$, which is asked to vanish for a {\em creatable} universe?

In a series of papers we and other authors \cite{NewCreatable,Lapiedra-Saez,Ferrando} have explained why Gauss coordinate systems could be candidates for  these preferred coordinate systems. Let us reproduce here some of the arguments used in these references.

To begin with, whatever be the complex used, $P^0$ is initially expressed as a $3$-volume integral, whose integrand is calculated at a given $t = t_0 = constant$, where $t$ is the time coordinate used. In other words, the different elementary contributions to this integral are all them calculated at the same time $t_0$. But, in order that this equal time has a physical significance, 
this time has to be a synchronized time, that is, one which gives the same reading for distant events which are physically simultaneous (we call this a {\em universal} time), this simultaneity being defined operationally like in Minkowski space-time (on this point we refer the reader to the book \cite{Landau},  epigraph $84$) . This {\em universal} character of time requires the use of coordinates such that the metric components $g_{0i}$ vanish.

On the other hand, if we want to define some sort of physical energy $P^0$, we should use as a time coordinate a physical (proper or canonical) time, that is we must have $g_{00} = -1$.

All in all, a good coordinate system in order to produce a sound physical energy $P^0$ would be a Gaussian one, at last at the 3-space infinity.

Furthermore, as it has been pointed in Section \ref{sec-2} and is well known, a Gauss coordinate system is one which is {\em adapted} to particles which fall freely in the space-time considered. In other words, their motion equations are $x^{i} = x^{i}_0$, where the   $x^{i}$ are the $3$-space coordinates  and the $x^{i}_0$ are three arbitrary constants corresponding to one of these free-falling particles. This {\em adapted} character of the Gauss coordinates says us that defining $P^0$ for an asymptotically Minkowskian space-time in these Gauss coordinates can be seen as a sort of generalization to General Relativity of what is called the proper energy of an $m_0$ mass particle in Minkowski space, which is just the particle energy seen by an instantaneously comoving inertial system, that is $m_0$. Actually, a suitable family of free-falling non rotating observers is the generalization, when gravitation is present, of an inertial coordinate system in $M_4$. Furthermore we will see next that our Schwarzschild metric has vanishing linear and angular $3$-momenta in the Gauss coordinates of 
(\ref{LemaitreP}), that is, these coordinates can also been seen as everywhere {\em comoving} coordinates which, by the way, are asymptotically at rest with respect to the $m$ mass. Thus, when $P^0$ is calculated in suitable Gaussian coordinates we could see this energy like a good generalization to General Relativity of the proper energy of a particle in the Minkowski space, and we could define this $P^{0}$ as the {\em proper} energy of the asymptotically Minkowskian space-time considered, as we have done in \cite{NewCreatable}.

The problem with this definition  in \cite{NewCreatable} is that it has now a natural competitor, for instance, in the case of the Schwarzschild metric: namely, when we take static standard coordinates (see (\ref{Smetrica})), since these coordinates are everywhere at rest with respect to the $m$ mass (our corresponding Gauss coordinates were at rest only asymptotically) and are Gauss coordinates though, differently to our everywhere Gauss coordinates, only asymptotically. Then, hereafter, we will reserve the name {\em proper} energy of a static asymptotically Minkowskian space-time to a $P^{0}$ calculated in asymptotically rectilinear coordinates everywhere at rest. Then, we will denote as {\em intrinsic} energy the $P^{0}$ calculated in suitable Gauss coordinates, at last in the present case of the Schwarzschild metric. 

The name {\em intrinsic} alludes to the fact that referring this metric to our Gaussian coordinates at rest at the infinity seems to add nothing to the metric itself, differently to the case of the coordinates associated to observers everywhere at rest, these observers needing virtually some non-gravitational action to avoid their otherwise unavoidable free fall. Furthermore, it is this {\em intrinsic} character what enable us to chose this {\em intrinsic} energy as the one which must vanish for a {\em self-creatable} ({\em creatable} for short) universe. Furthermore, it is this kind of {\em intrinsic} energy that vanishes for a closed or flat Friedmann-Lema\^{\i}tre-Robertson-Walker (FLRW) universe according to most literature on the subject (cf. \cite{NewCreatable,Rosen,Cooperstook}).

All in all, the reason for two sound different values for the Schwarzschild metric energy, the {\em proper} energy, $P^{0} = m$, and the {\em intrinsic} one,  $P^{0} = 0$, is simply the use of different coordinates with different operational meanings. In the first case, the use of coordinates associated everywhere to at rest observers, these coordinates becoming asymptotically Gauss coordinates. In the second case, the use of everywhere Gauss coordinates, which become associated to asymptotically at rest observers.

Finally, let us consider the following uniqueness question: given an asymptotically Minkowskian metric, referred to some Gauss coordinate system, 
does $P^0$ depend on the particular asymptotic Gauss system considered? The answer to this question is positive: aside the double result obtained in the Schwarzschild case, $P^0 = m$
or $P^0 = 0$, it is enough to remember that, even a mere Lorentz transformation 
(any boost, actually) at the infinity $r \to \infty$, will change the $P^0$ value (see \cite{Weinberg}, chapter 7, epigraph 6, if necessary). But, then, which, if any, are the good Gauss coordinate systems to be used in order to calculate an {\em intrinsic} energy?

This question will be partially addressed in the next section where we consider the {\em creativeness} (see \cite{NewCreatable} and references therein) of the Schwarzschild space-time.


\section{Creativeness of the Schwarzschild space-time}
\label{sec-6}

The same question raised at the end of the precedent section, can be raised in the more general case of NON asymptotically Minkowskian space-times. The question has been treated in three papers \cite{Ferrando,Lapiedra-Saez,NewCreatable} from us and other authors. The partial answer was the following one: when we deal with a {\em universe} (i.e., a space-time whose well defined linear $4$-momentum and angular $4$-momentum are conserved), in order to define its intrinsic momenta, we must use coordinates such that:

(a) be Gaussian coordinates, 

(b) be such that  the $3$-space metric is manifestly conformally flat on the $2$-surface boundary of the space-like $3$-surface $T = T_0$ (that is, for $r \to \infty$, for $T = T_0$), 

(c) be such that the corresponding linear $3$-mo\-men\-tum, $P^{i}$, and angular $3$-momentum, $J^{ij}$, vanish, the last one irrespective of its origin. 

Such coordinate systems can be proved to exist for any universe, and we call them {\em intrinsic} coordinate systems. But, it is still possible that 
different intrinsic coordinate systems exist for the same universe, leading perhaps to different $4$-momenta. Nevertheless, if we find one of these intrinsic coordinate systems 
such that the corresponding $4$-momenta vanish, we must conclude that the proper $4$-momenta of this universe vanish in themselves, and then we call it a {\em creatable} universe. We must conclude this by noticing that Minkowski space is trivially creatable in this precise sense, even if it can be shown that there are intrinsic coordinate systems for which its $4$-momenta do not all them vanish \cite{NewCreatable}. Thus, if a {\em creatable} universe has intrinsic coordinate systems whose corresponding $4$-momenta do not completely vanish, we must interpret that this is due 
to the fact that these intrinsic coordinate systems  fail to respect some symmetries of the corresponding universe. Which of them? Just the ones that allow  us to find coordinate 
systems where these two $4$-momenta, the linear and the angular ones, vanish. The name {\em creatable} universe comes from the suggestion that we 
need this vanishing in order that this universe could arise from a vacuum quantum fluctuation \cite{Albrow,Tryon}.

Applying these ideas to the metric (\ref{LemaitreP}) of a black hole, or of a Schwarzschild  metric whose source radius is larger than $r_0$, 
we first see that because of the spherical symmetry, the above $P^{i}$ and  $J^{ij}$ must vanish. This means that coordinates $(T, \rho, \theta, \phi)$ in (\ref{LemaitreP})
satisfy the above property (c). Furthermore, these coordinates are Gaussian coordinates (property (a)), while property (b) is satisfied  {\em a fortiori} since for $r \to \infty$ and $T = T_0$
the corresponding $3$-space metric is just manifestly flat. All in all,  $(T, \rho, \theta, \phi)$ are an example of intrinsic coordinates 
according to the definition just given. But the same spherical symmetry tells also us that $J^{0i}$, the mixed components of the angular $4$-momentum,  
vanish too,  and since we have seen that the {\em intrinsic} energy,  vanishes, 
we must conclude that any non-rotating black hole, or Schwarzschild space with a source radius larger than $r_0$, are both {\em creatable} universes. 

On the other hand, a closed FLRW universe, perturbed or not,  is a {\em creatable} universe \cite{Lapiedra-Saez}. However, from what has been concluded in the present paper, 
it is not clear whether a perturbed  non-rotating black hole, i.e., an slightly rotating one, would still be a {\em creatable} universe. In other words, the creativeness of a non-rotating black hole could be a non-stable result. Further on, we could ask if some black hole like metrics \cite{Vir-Parikh} would have a vanishing {\em intrinsic} energy, and if this hypothetical result would be stable.


\section{Final considerations: why the Weinberg complex?}
\label{sec-7}
 
 Let us consider some complex, $\tau^{\alpha\beta}$, in General Relativity, that is $\tau^{\alpha\beta} = \eta^{\alpha\mu} \eta^{\beta\nu} (T_{\mu\nu} + t_{\mu\nu})$, 
  with  $T_{\alpha\beta}$ the energy--momentum tensor, $\eta^{\mu\nu}$ the Minkowski tensor and $t_{\alpha\beta}$ some ``pseudo--tensor" associated 
  to the presence of the gravitational field, such that the following continuity equation
\begin{equation}\label{divtau}
\partial_\alpha \tau^{\alpha\beta} = 0
\end{equation}
becomes true. From (\ref{divtau}), the following balance equation
\begin{equation}\label{balanceE1}
\frac{d}{dx^{0}} \int_V  \tau^{\alpha 0} d x^3 = -  \int_{\Sigma(V)}  \tau^{\alpha i} d \Sigma_i, 
\end{equation}
and in particular,  the relation
\begin{equation}\label{balanceE2}
\frac{d}{dx^{0}} \int_V  \tau^{00} d x^3 = -  \int_{\Sigma(V)}  \tau^{0i} d \Sigma_i, 
\end{equation}
follow, the last equation giving the balance between the variation in time of the energy enclosed in a given $3$-volume, $V$,  
and the flux of this energy trough the boundary $2$-surface, $\Sigma(V)$, of this volume.

Given some complex $\tau^{\alpha\beta}$, we can add to it any arbitrary quantity $\partial_\gamma h^{\gamma \alpha \beta}$, such that $ h^{\gamma\alpha\beta}= - h^{\alpha\gamma\beta}$, to get another complex
\begin{equation}\label{tau-tilde}
\tilde{\tau}^{\alpha\beta}  = \tau^{\alpha\beta} + \partial_\gamma h^{\gamma \alpha \beta}, 
\end{equation}
satisfying trivially its own continuity equation
\begin{equation}\label{divtautilde}
\partial_\alpha \tilde{\tau}^{\alpha\beta} = 0, 
\end{equation}
leading to the new balance relation
\begin{equation}\label{balanceE1-tilde}
\frac{d}{dx^{0}} \int_V  \tilde{\tau}^{\alpha 0} d x^3 = -  \int_{\Sigma(V)}  \tilde{\tau}^{\alpha i} d \Sigma_i, 
\end{equation}
as much valid as the original one (\ref{balanceE1}). If we focus our attention on this balance, (\ref{balanceE1}) and (\ref{balanceE1-tilde}) are on the same foot, but if 
what we want is, for example, to calculate $P^0$, the energy of the corresponding space-time, we are going to obtain, in general, different values for this energy, 
$\int \tau^{00} d^3x$, or $\int \tilde{\tau}^{00} d^3x$, according to what complex we choose. Which one, if any, should we chose?  

In the above quoted reference \cite{AguirreAV-96}, it is proved that the Einstein, Landau-Lifshitz, Tolman, Papapetrou, and Weinberg complexes, have (essentially) the same expression for the Kerr-Newman family of metrics, in Kerr-Schild  ``Cartesian coordinates". Furthermore, the same energy density is found for the Vaidya \cite{Vir-Pramana-92} and Einstein-Rosen metrics 
\cite{Rosen-Vir-93,Vir-Pramana-95} using the different Einstein, Tolman and Landau-Lifshitz complexes.

Nevertheless in \cite{Vir-91} different energy densities are found for some of these complexes in the case of some particular metrics of the Kerr-Newman family in another coordinates. Also the results from M{\o}ller complex are not always in accordance from those coming from the other quoted complexes \cite{Vir-90ab}. 

All this means that the question of what complex, if any, should be chosen, in general, or in a particular situation, is a valid one, our answering being 
that the Weinberg complex is a specially good candidate for such election because it comes {\em directly} from the cornerstone of the General Relativity 
building, i.e., the Einstein field equations. By the word {\em directly} we mean that the Weinberg complex appears in  
a (non manifestly covariant) way of writing the Einstein field equations by merely reordering its different terms between both hand sides,  
without adding any term like $\partial_\gamma h^{\gamma \alpha \beta}$ \cite{Weinberg}. Furthermore, in the Introduction we have recalled
that the energy $P^0$ given by ($\ref{energy-S}$), from the Weinberg complex, is the same as the Arnowitt-Deser-Misner (ADM) energy \cite{ADM}. 
We could then make the conjecture that this is so because the ADM energy comes again from the Einstein field equations, this time written in the standard 
$3+1$ formalism of the General Relativity 
(see \cite{EricG} for an extensive account). In all, this is why, in the present paper, we have used the Weinberg complex in such a preferential way.

On the other hand,  the Newtonian gravitational energy per unit mass, $\varepsilon$,  is defined from the Newtonian potential $\varphi$ up to an arbitrary additive constant $C$, such that 
$\varepsilon = \varphi + C$. It can be easily seen that this arbitrary constant generates a correction in the post--Newtonian terms of the metric which are by no means deprived of physical effects. Furthermore, let it be the different post--Newtonian solutions to Einstein field equations associated with the different $C$ values. Impose suitable physical conditions to these solutions in absence of gravitational radiation, i.e., for $r \to \infty$, the metric becomes manifestly flat (remember, on the other hand,  that, we can guarantee the existence of a unique $P^0 \neq 0$ by making sure
that the metric approaches the Minkowski metric as  $r^{-1}$ when $r \to \infty$). In this way we select a unique physical solution of the Einstein field equations to which corresponds the value $C=0$. Which, again suggests that the above selection of the Weinberg complex, with its {\em direct} relation to this equations, could be a consistent one.

\vspace{1cm}


{\bf Acknowledgements} We thank professors M. Portilla and D. S\'{a}ez for their criticisms and comments on the manuscript. 
This work has been supported by the Spanish
ministries of ``Ciencia e Innovaci\'on'' and ``Econom\'{\i}a y Competitividad'' 
MICINN-FEDER projects FIS2009-07705 and FIS2012-33582.

\appendix
\section{Some remarks on a result by Weinberg}
\label{ap-A}

In his book \cite{Weinberg}, epigraph $6$ (``Energy, momentum and angular momentum of gravitation"), chapter $7$, Weinberg referring to the linear 
$4$-momentum, $P^\alpha$, of the gravitational field and its sources, states that $P^\alpha$ ``have the important property of being invariant under any coordinate transformation that reduces
at infinity to the identity". Then he writes such a transformation as
\begin{equation}\label{W-trans}
x'^\mu = x^\mu + \epsilon^\mu(x)
\end{equation}
``where  $\epsilon^\mu(x)$ vanishes as $r \to \infty$".

To prove it, Weinberg writes $P^\lambda$ as
\begin{equation}\label{W-linear-momentum}
P^\lambda = - \frac{1}{8\pi} \lim_{r \to \infty} \int Q^{i0\lambda} n_i r^2 d \Omega,
\end{equation}
where quantities $Q^{i0\lambda}$ depend on the space and time derivatives of $h_{\mu\nu} = g_{\mu\nu} - \eta_{\mu\nu}$, and where $\eta_{\mu\nu}$ is the Minkowski tensor. In particular, for $P^0$
we have (\ref{energy-S}). Then, the coordinate change (\ref{W-trans}) induces an elementary change $\Delta Q^{i0\lambda}$ which can be written as
\begin{equation}\label{W-increment}
\Delta Q^{i0\lambda} = \partial_j D^{ji0\lambda}, 
\end{equation}
$D^{ji0\lambda}$ being a linear combination of products of some components of $\partial_\gamma \epsilon^\mu$ and $\eta^{\lambda\rho}$ (see \cite{Weinberg} for details), $D^{ji0\lambda}$ 
being antisymmetric in its $j, i$ indices. This elementary change $\Delta Q^{i0\lambda}$ entails the corresponding elementary change $\Delta P^\lambda$
\begin{equation}\label{W-delta-linear-momentum}
\Delta P^\lambda = - \frac{1}{8\pi} \lim_{r \to \infty} \int \partial_j D^{ji0\lambda} n_i r^2 d \Omega,
\end{equation}
that applying Gauss theorem (assuming that  the corresponding $3$-volume integrand is a continuous function) gives
\begin{equation}\label{W-delta-lm-zero}
\Delta P^\lambda = - \frac{1}{8\pi}  \int \partial_i \partial_j D^{ji0\lambda} d^3x = 0,
\end{equation}
since $D^{ji0\lambda} = - D^{ij0\lambda}$, as the author wanted to prove. In particular $\Delta P^0 = 0$.

The first remark to be made about this reasoning is that we must precise the exact behavior of the above coordinate transformation
``that reduce at infinity to identity''. To begin with, imagine that matrix $\partial_\gamma \epsilon^\mu$ goes at infinity to zero as the $h_{\mu\nu}$ go. Then, if for $r \to \infty$, 
$h_{\mu\nu} \to 0$ slower than $r^{-1}$, integrals (\ref{W-linear-momentum}) and  (\ref{W-delta-linear-momentum}) would diverge in general, while with $h_{\mu\nu} \to 0$ fast than $r^{-1}$, both integrals vanish, and only if $h_{\mu\nu} \to 0$ just as $r^{-1}$ we obtain two finite results.

But, what happens when, as it is the case in our Section \ref{sec-4}, the transformed components, $h'_{\mu\nu}$, go like $r^{-3/2}$, while the original ones, 
$h_{\mu\nu}$, go like $r^{-1}$? This means that the corresponding 
matrix $\partial_\gamma \epsilon^\mu$ has terms which go like $r^{-1}$ and other ones which go like $r^{-3/2}$, such that finally $h_{\mu\nu}$ and $h'_{\mu\nu}$ behave differently at infinity: as
$r^{-1}$ and $r^{-3/2}$, respectively. This double behavior leads to the different results $P^0 = m$ and $P'^0 = 0$ in Section \ref{sec-4}, showing how in this case the Weinberg proof works.

In all, for asymptotically Minkowskian metrics, referred to different asymptotic rectilinear coordinates, we can have different $P^0$ values.

\section{Proving the flat-statement}
\label{ap-B}

For the sake of completeness, let us consider here a simple proof of the property announced in the text: 

{\em Any space-time metric of the form%
\begin{equation}\label{metrica}
 ds^2= - dT^2 +  B(R, T) [dR^2 + R^2(d \theta^2 + \sin^2 {\hspace{-0.7mm}} \theta \, d \phi^2)] , 
\end{equation}
that is a vacuum solution of the Einstein field equations, is necessarily a locally flat metric.}

By using the standard $3+1$ formalism of the General Relativity (see, for example, \cite{EricG}), the notions involved in the proof 
become very transparent.

The vector field $u = \partial_T$ defines a free-falling radial congruence of observers which, in addition,  is vorticity and shear-free. 
The extrinsic curvature ${\cal K}$ of the slices $T = {\rm constant}$ 
is proportional to the induced metric $\gamma$ on these slices, and it is determined by a sole function $\Phi$, 
 \begin{equation}\label{Phi}
 \Phi = - \frac {\dot{B}}{2 B}
\end{equation}
where the `dot' stands for the partial derivative with respect to $T$. 

In fact, ${\cal K}^R_R = {\cal K}^\theta_\theta = {\cal K}^\phi_\phi = \Phi$, are the sole non identically vanishing components of ${\cal K}$, and then,  
function $\Phi$ is related to the expansion of $u$, 
\begin{equation}\label{3Phi}
\nabla_\mu u^\mu = - {\cal K}^\mu_\mu = - 3 \Phi.
\end{equation}

For a vacuum metric form (\ref{metrica}),  the energy flux vanishes and the momentum constraint says that $\Phi$ does not depend of $R$, $\Phi (T)$. The 
remaining Einstein equations are written as (see, for instance, reference \cite{Alicia}):

\begin{equation}\label{C1}
3 \Phi^2 = - \frac{{\cal R}}{2}, 
\end{equation}
\begin{equation}\label{E1}
(\Phi B)^{\cdot} =  B  \frac{{\cal R}}{2} - \frac{F}{R^2} + B \Phi^2, 
\end{equation}
 
 \begin{equation}\label{E2}
(\Phi B)^{\cdot} =  B \frac{{\cal R}}{4} + \frac{F}{2 R^2} + B \Phi^2, 
\end{equation}
${\cal R}$ being the scalar curvature of  $\gamma$,  and $F$ is the function
 \begin{equation}\label{F}
 F = - \frac{RB'}{4 B^2} (R B' + 4 B).
\end{equation}
where the  `prime' stands for the partial derivative with respect to $R$. 

Taking into account the constraint equation  (\ref{C1}) (energy constraint), the evolution equations (\ref{E1}) and (\ref{E2}) are written as:
\begin{equation}\label{E1-2}
(\Phi)^{\cdot} =   \Phi^2, 
\end{equation}
 and
 \begin{equation}\label{E2-2}
 F = - B R^2 \Phi^2.
\end{equation}

The integration is easily accomplished by considering separately the cases $\Phi = 0$ and  $\Phi \neq 0$.

(i) If  $\Phi = 0$, Eq. (\ref{Phi}) says that $B$ does not depend of $T$, $B(R)$, and from (\ref{E2-2}),  $F=0$. Then, from (\ref{F}), $B={\rm constant}$ or $B R^4= {\rm constant}$, 
and the metric form (\ref{metrica}) becomes the Minkowski metric.  The last solution is mapped 
in the standard Minkowski form by performing a radial inversion,  $R^* = 1/R$.

(ii) If  $\Phi \neq 0$, Eqs. (\ref{Phi}) and (\ref{E1-2}) lead to
\begin{equation}\label{PhiB}
\Phi = -\frac{1}{T+a},  \quad B = f(R) (T + a)^2, 
\end{equation}
with $a$ an arbitrary constant and $f(R)$ obeying the differential equation 
 \begin{equation}\label{eq-de-f}
f' (R f' + 4f) = 4 f^3 R
\end{equation}
as it  follows from making  compatible Eqs. (\ref{F}) and (\ref{E2-2}). This equation may be conveniently written as 
 \begin{equation}\label{eq-de-f-fac}
\big(f' +  \frac{2f}{R}\big)^2  - \frac{4 f^2}{R^2} (1 + R^2 f) = 0, 
\end{equation}
and then, it is easy to see that the  general solution is given by 
 \begin{equation}\label{f}
f(R) = \frac{4 b^2}{(1-b^2 R^2)^2}, \quad R \in (0, 1/b) \cup (1/b,  \infty)
\end{equation}
$b$ being an arbitrary constant. Constants $a$ and $b$ are non-essential because they may be absorbed by a trivial redefinition of the coordinates, 
$\tau = T + a$ and $r = 2 b R$. 

Therefore, when $\Phi\neq 0$, the sole metric form (\ref{metrica}) that is a solution of the vacuum Einstein equations is the Milne metric:
\begin{equation}\label{Milne}
 ds^2= - d\tau^2 + \frac {\tau^2}{(1 - \frac{r^2}{4})^2}  [dr^2 + r^2(d \theta^2 + \sin^2 {\hspace{-0.7mm}} \theta \, d \phi^2)] 
\end{equation}
that describes a locally flat (Minkowskian) expanding universe. 

Notice that the radial inversion $r \to 4/r$ isometrically 
maps the regions $r<2$ and $r>2$ each other.

%
%
%


\bibliography{apssamp}

\end{document}